\documentclass[floatfix]{revtex4}
\usepackage{blindtext}
\usepackage[export]{adjustbox}
\usepackage{graphicx,epsfig,epstopdf}
\usepackage{amsmath}
\usepackage{color}
\usepackage{subfigure}
\usepackage{array}
\usepackage{tabularx}
\usepackage{multirow}
\newcolumntype{L}[1]{>{\raggedright\arraybackslash}p{#1}}
\newcolumntype{C}[1]{>{\centering\arraybackslash}p{#1}}
\newcolumntype{M}[1]{>{\centering\arraybackslash}m{#1}}

\def\_#1{{\bf #1}}
\def\.{\cdot}

\begin{document}

	\title{Non-reciprocal sound propagation in space-time modulated media}
	
	\author{Junfei Li}
	\email{junfei.li@duke.edu}
	\affiliation{ 
		Department of Electrical and Computer Engineering, Duke University, Durham, North Carolina 27708, USA
	}
	
	\author{Chen Shen}
	
	\affiliation{ 
		Department of Electrical and Computer Engineering, Duke University, Durham, North Carolina 27708, USA
	}
	
	
	
	\author{Yangbo Xie}
	\affiliation{ 
		Department of Electrical and Computer Engineering, Duke University, Durham, North Carolina 27708, USA
	}%
	
	\author{Steven A. Cummer}%
	\email{cummer@ee.duke.edu}
	\affiliation{ 
		Department of Electrical and Computer Engineering, Duke University, Durham, North Carolina 27708, USA
	}

	\begin{abstract}
		Realization of non-reciprocal devices, such as isolators and circulators, is of fundamental importance in microwave and photonic communication systems. This can be achieved by breaking time-reversal symmetry in the system or exploiting nonlinearity and topological effects. However, exploration of non-reciprocal devices remains scarce in acoustic systems. In this work, sound propagation in a space-time modulated medium is theoretically studied. Finite-difference time-domain (FDTD) simulations are carried out to verify the results. Functionalities such as mode conversion, parametric amplification and phase conjugation are demonstrated. 
	\end{abstract}

	\maketitle

	\section{Introduction}
	
	Reciprocity is a fundamental principle for most wave systems in which the relationship between a source at one point and the measured response at another point is symmetric when the source and measurement points are interchanged. For a long time, non-reciprocal devices that break this symmetry have been pursued since there are many practical situations where breaking reciprocity can be advantageous. In electric circuits, non-reciprocity can be easily achieved through nonlinear semiconductor devices, and diodes and transistors are widely used in almost all electronic systems. For optics and electromagnetics, non-reciprocity has been demonstrated using magnetic field biasing \cite{adam2002ferrite}, nonlinearity \cite{sounas2017time,sounas2018broadband}, systems with angular momentum bias \cite{sounas2013giant} and topological insulators \cite{ rechtsman2013photonic, lu2014topological,khanikaev2017two}. Despite the growing interest for non-reciprocity in optics and electromagnetics in recent years, non-reciprocal phenomena and devices for acoustic waves are less explored \cite{fleury2015nonreciprocal}. 
	
	For acoustic systems, non-reciprocity has been primarily achieved through nonlinear effects that partially convert the energy in fundamental modes into higher harmonics, and then applying spatial or temporal frequency filters \cite{liang2009acoustic,liang2010an,boechler2011bifurcation,popa2014non-reciprocal}. However, nonlinear effects are weak for most materials, and the amplitude needs to be impractically high to induce significant nonlinear effects. Also, it is difficult to have full control over the spectrum through nonlinear effects because of the generation of many harmonics in these systems. Another approach is to introduce directional bias by constant air flow \cite{fleury2014sound}. However, this approach requires mechanical motion that is associated with high energy consumption, which can create challenges for device-based implementations. In recent years, topological insulators that break reciprocity with external bias, such as constant flow or directional modulation, have also attracted much attention \cite{yang2015topological,ni2015topologically,fleury2016floquet, zhu2018experimental}. However, such systems require sophisticated control of flow field or multiple coupled resonances who are sensitive to losses, making them hard to implement in experiments. 
	
	Breaking time-reversal symmetry using space-time modulation has been studied in time-varying transmission lines for many decades \cite{tien1958traveling,tien1958parametric,cullen1958travelling,qin2014nonreciprocal}. Recently, non-reciprocity through space-time modulation has returned to the spotlight and the idea has been applied to modern optical and electromagnetic systems and metasurfaces \cite{poulton2012design,hafezi2012optomechanically, sounas2014angular,hadad2015space, correas2016nonreciprocal, ruesink2016nonreciprocity,hadad2016breaking,taravati2017mixer, miri2017optical}. For mechanical waves, space-time modulated elastic beams have been proposed to create a directional band gap \cite{trainiti2016non}. Space-time modulated waves in mass-spring systems have also been proposed to create directional wave manipulation for elastic waves \cite{nassar2017non,wang2018observation}. For acoustics, time modulated circulators have been studied using coupled mode theory \cite{fleury2015subwavelength,koutserimpas2018coupled}.
	
	In this paper, acoustic wave propagation in a general space-time modulated medium is studied by directly solving the time-varying wave equations instead of using time Floquet theory to analyze its bandgaps. We follow an approach originally applied to time-varying transmission line system \cite{tien1958parametric}.  For such a system, two otherwise orthogonal waveguide modes can be coupled efficiently under certain conditions, giving rise to inter band and intra band phonon transition. Different from using time Floquet theory, directly solving the equations provides detailed information on how do waves change gradually in such a system. By applying different types of modulation, different non-reciprocal functionalities can be achieved including one-way frequency conversion, parametric amplification, and phase conjugation.  Modulation conditions under which each of these behaviors emerge are derived, and finite difference time domain (FDTD) simulations of the full dynamical system show good agreement with the theoretical predictions. The proposed time-reversal breaking system can find applications in many aspects that allow unprecedented wave control capability. For example, frequency conversion enables directional band gap for acoustic waves, rectifiers, advanced spectrum control for communication and energy transmission; parametric amplification provides a possible implementation for the gain media in parity-time (P-T) symmetric systems and powerful speaker designs; and phase conjugation can be useful for an all-angle retro-reflector, acoustic lasing, data processing and acoustic communications.

	\section{Non-reciprocal sound propagation through space-time modulated media}
	
	In this section, we will start from solving the space-time varying wave equation and investigate the acoustic wave propagation in a medium whose density is varying with both space and time. Similar coupled wave solution can be found in \cite{tien1958traveling} where two transmission lines are coupled with varying inductors, while in our paper we focus on acoustic wave propagation in one waveguide. We will show that by applying specific types of modulation, two otherwise orthogonal waves will be coupled and phenomena like frequency conversion and parametric amplification will arise. Then we will show that these phenomena will also arise when the bulk modulus is modulated the same way. Here we assume the wave amplitude is small so that nonlinear effects are negligible.
	
	\subsection{Sound propagation in media with space-time modulated density}
	
	\begin{figure}
		\includegraphics[width=0.7\linewidth]{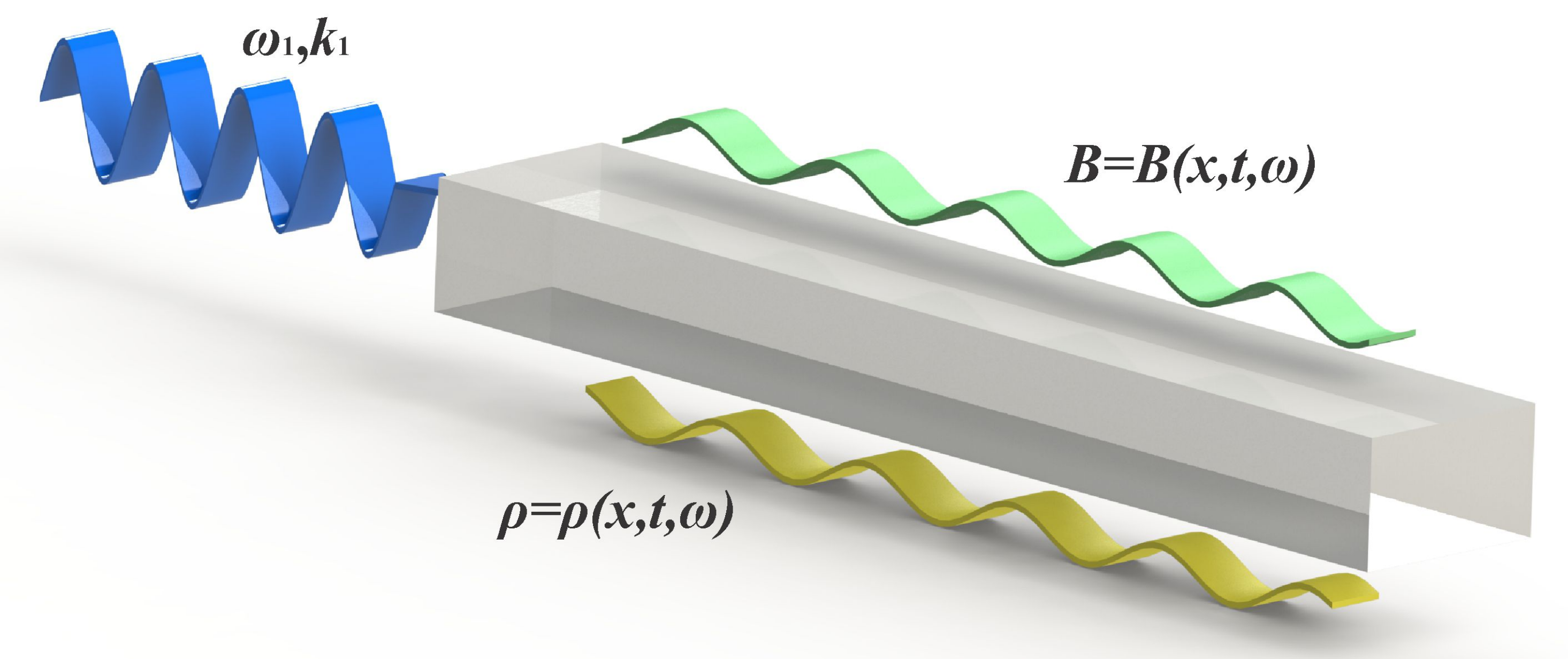}
		\caption{Spatial-temporally modulated medium under study. The effective density or effective compressibility is dependent on time, space and frequency.} 
		\label{fig:Fig7}\end{figure}
	
	We begin by considering a one-dimensional waveguide in which the effective density $\rho$ of the medium is modulated in both space and time while bulk modulus remains constant. Generally $\rho$ is dispersive, i.e. $\rho=\rho(x,t,\omega)$. According to Newton's second law and Hooke's law,
	
	\begin{equation}
		\begin{aligned}
			-\frac{\partial p}{\partial x}=\rho\frac{\partial v}{\partial t}\\
			-\frac{\partial p}{\partial t}=\kappa\frac{\partial v}{\partial x}
		\end{aligned}
		\label{eq:NewtonHooke1}\end{equation}
	
	Taking partial derivative with respect to $t$ and $x$ respectively, we get
	\begin{equation}
		\begin{aligned}
			-\frac{\partial^2p}{\partial x \partial t}=\frac{\partial \rho}{\partial t}\frac{\partial v}{\partial t}+\rho\frac{\partial^2v}{\partial t^2}\\
			-\frac{\partial^2p}{\partial x \partial t} = \kappa\frac{\partial^2v}{\partial x^2}
		\end{aligned}
		\label{eq:NewtonHooke1Partial}\end{equation}
	
	Combining the two equations to eliminate $p$ we can get the wave equation in velocity, namely
	
	\begin{equation}
		\frac{\partial^2v}{\partial x^2}=\frac{1}{\kappa}\frac{\partial \rho}{\partial t}\frac{\partial v}{\partial t}+\frac{\rho}{\kappa}\frac{\partial^2v}{\partial t^2}
		\label{eq:WaveEqnRho}
	\end{equation}
	
	Let us assume the solution is composed of two waves with different frequencies in the waveguide with the form 
	
	\begin{equation}
		\begin{aligned}
			v_1=A_1(x)e^{\mathrm{j}(\omega_1t-k_1x)}\\
			v_2=A_2(x)e^{\mathrm{j}(\omega_2t-k_2x)}
		\end{aligned}
		\label{eq:TwoWavesv}\end{equation}
	
	where $k_1=\frac{\omega_1}{c_1}$, $k_2=\frac{\omega_2}{c_2}$, and $c_1=\sqrt{\frac{\kappa}{\rho_1}}$, $c_2=\sqrt{\frac{\kappa}{\rho_2}}$.   The total wave velocity is given by $v=v_1+v_2$.
	
	Let us further assume that the density modulation is sufficiently weak that $A_1$ and $A_2$ are slowly varying and $\frac{\partial^2A_{1,2}}{\partial x^2}$ are negligible.  We assume the density variation is described by
	
	\begin{equation}
		\rho=\rho_1[1+m_1\cos(\Omega t-\beta x)]
		\label{eq:RhoChange1}\end{equation}
	for the wave with frequency $\omega_1$, and 
	\begin{equation}
		\rho=\rho_2[1+m_2\cos(\Omega t-\beta x)]
		\label{eq:RhoChange2}\end{equation}
	for the wave with frequency $\omega_2$.
	
	Typically, density is just a property for a material. However for metamaterials, the effective density is controlled by the parameters of the metamaterial structures, such as the mass and in-plane tension for membranes, and the effective density is usually dispersive.  Therefore, the modulation depth for different frequencies can be different. The quantities $m_1$ and $m_2$ represent the amplitude of effective density change when the controlling parameter changes and it follows $m_1=m_2$ in a nondispersive medium.
	
	\subsection{Unidirectional frequency conversion}
	
	The general space-time modulation described above admits a number of different types of solution.  We focus first on unidirectional frequency conversion.  Here we define 
	\begin{equation}
		\begin{aligned}
			\Omega=\omega_1-\omega_2 \\
			\beta=k_1-k_2
		\end{aligned}
		\label{eq:AssumeConversion}\end{equation}
	
	Putting these into Equation
	(\ref{eq:WaveEqnRho}), we find that the two frequencies are coupled through the time-varying density $\rho(x,t,\omega)$. It can be easily shown that 
	
	\begin{equation}
		\begin{aligned}
			B_1e^{\mathrm{j}(\omega_1t+k_1x)}; B_2e^{\mathrm{j}(\omega_2t+k_2x)} \\
			A_1^*e^{-\mathrm{j}(\omega_1t-k_1x)}; A_2^*e^{-\mathrm{j}(\omega_2t-k_2x)} \\
			B_1^*e^{-\mathrm{j}(\omega_1t+k_1x)}; B_2^*e^{-\mathrm{j}(\omega_2t+k_2x)}
		\end{aligned}
		\label{eq:CoupledPairs}\end{equation}
	are also coupled solutions. $A_1e^{\mathrm{j}(\omega_1t-k_1x)}$ is a forward traveling wave and $B_1e^{\mathrm{j}(\omega_1t+k_1x)}$ is a backward traveling wave, and $A_1^*e^{-\mathrm{j}(\omega_1t-k_1x)}$ and $B_1^*e^{-\mathrm{j}(\omega_1t+k_1x)}$ are their complex conjugates. The complete solution of Equation
	(\ref{eq:WaveEqnRho}) is composed of four mixed pairs. Without loss of generality, here for illustration we just analyze the first coupled pair. We put the form of assumed waves and prescribed modulation into Equation
	(\ref{eq:WaveEqnRho}),  
	neglect the $\frac{\partial^2A_{1,2}}{\partial x^2}$ terms and equate the terms with the same frequency.  The result is the following two equations
	
	\begin{equation}
		\begin{aligned}
			\frac{\partial A_1}{\partial x}=-\mathrm{j} \frac{\rho_2m_2}{4\kappa}\frac{\omega_1\omega_2}{k_1}A_2\\
			\frac{\partial A_2}{\partial x}=-\mathrm{j} \frac{\rho_1m_1}{4\kappa}\frac{\omega_1\omega_2}{k_2}A_1
		\end{aligned}
		\label{eq:EquTwoAmplitudes}\end{equation}
	
	By taking partial derivative and eliminating $A_2$, we get a partial differential equation for $A_1$:
	
	\begin{equation}
		\frac{\partial^2A_1}{\partial x^2}=-\frac{m_1m_2k_1k_2}{16}A_1
		\label{eq:EqnA1}\end{equation}
	
	Solving this equation yields
	
	\begin{equation}
		A_1=a_1e^{\mathrm{j}\alpha x}+b_1e^{-\mathrm{j}\alpha x}
		\label{eq:FormA1}\end{equation}
	
	where $\alpha=\frac{\sqrt{m_1m_2k_1k_2}}{4}$ and $a_1$ and $b_1$ are constants decided by the boundary conditions. Putting the form of $A_1$ into Eqn.(\ref{eq:EquTwoAmplitudes}) to solve $A_2$, we get
	
	\begin{equation}
		A_2=-\sqrt{\frac{m_1k_1\rho_1}{m_2k_2\rho_2}}(a_1e^{\mathrm{j}\alpha x}-b_1e^{-\mathrm{j}\alpha x})
		\label{eq:FormA2}\end{equation}
	
	Therefore, we finally have the solution 
	
	\begin{equation}
		\begin{aligned}
			v_1=(a_1e^{\mathrm{j}\alpha x}+b_1e^{-\mathrm{j}\alpha x})e^{\mathrm{j}(\omega_1t-k_1x)}\\
			v_2=-\sqrt{\frac{m_1k_1\rho_1}{m_2k_2\rho_2}}(a_1e^{\mathrm{j}\alpha x}-b_1e^{-\mathrm{j}\alpha x})e^{\mathrm{j}(\omega_2t-k_2x)}
		\end{aligned}
		\label{eq:SolutionvConvert}\end{equation}
	
	Similar results can be found with other coupled pairs. If at $x=0$, 
	
	\begin{equation}
		\begin{aligned}
			v_1(0,t)=a e^{\mathrm{j}(\omega_1t+\theta)}\\
			v_2(0,t)=0
		\end{aligned}
		\label{eq:TwoBCsv}\end{equation}
	
	the complete solution to equation(\ref{eq:WaveEqnRho}) is
	
	\begin{equation}
		\begin{aligned}
			v_1=a\cos(\alpha x)e^{\mathrm{j}(\omega_1t-k_1x+\theta)}\\
			v_2=-\sqrt{\frac{m_1k_1\rho_1}{m_2k_2\rho_2}}a\sin(\alpha x)e^{\mathrm{j}(\omega_2t-k_2x+\theta+\frac{\pi}{2})}
		\end{aligned}
		\label{eq:FinalSolutionvConvert}\end{equation}
	
	Therefore, the intensities of these two waves are
	
	\begin{equation}
		\begin{aligned}
			I_1=\frac{1}{2}\rho_1c_1|v_1|^2=\frac{1}{2}Z_1a^2\cos^2(\alpha x)\\
			I_2=\frac{1}{2}\rho_2c_2|v_2|^2=\frac{1}{2}Z_2\frac{m_1k_1\rho_1}{m_2k_2\rho_2}a^2\sin^2(\alpha x)
		\end{aligned}
		\label{eq:Intensities_1}\end{equation}
	
	where $Z_1=\rho_1c_1$ and $Z_2=\rho_2c_2$ are characteristic impedances of the two waves. Eqn.(\ref{eq:Intensities_1}) show that if we apply a signal of frequency $\omega_1$ at the input end of the waveguide, the power at that frequency is completely converted to that of $\omega_2$ in a distance of $\alpha x=\frac{\pi}{2}$. In the next segment of length $\alpha x=\frac{\pi}{2}$, power of frequency $\omega_2$ reverts to that of $\omega_1$, and then converts back and forth. At $\alpha x=0, \pi, 2\pi,...$, $|I_1|$ is at maximum, and at $\alpha x=\pi/2, 3\pi/2, 5\pi/2,...$, $|I_2|$ is at maximum. On the other hand, the backward propagating wave will not be affected since the generated mode is not supported in such a system. 
	
	\subsection{Unidirectional parametric amplification and phase conjugation}
	
	Another class of solution enabled by space-time modulation is parametric amplification, where the incident wave is amplified exponentially while propagating in such media. We begin by defining
	\begin{equation}
		\begin{aligned}
			\Omega=\omega_1+\omega_2 \\
			\beta=k_1+k_2
		\end{aligned}
		\label{eq:AssumeAmp}\end{equation}
	
	then the two coupled waves become
	
	\begin{equation}
		\begin{aligned}
			v_1=A_1(x)e^{\mathrm{j}(\omega_1t-k_1x)}\\
			v_2=A_2(x)e^{-\mathrm{j}(\omega_2t-k_2x)}
		\end{aligned}
		\label{eq:TwoWavesAmp}\end{equation}
	
	Inserting these into Equation
	(\ref{eq:WaveEqnRho}) and going through the similar process, we can find the governing equation for $A_1$ becomes 
	
	\begin{equation}
		\frac{\partial^2A_1}{\partial x^2}=\frac{m_1m_2k_1k_2}{16}A_1
		\label{eq:EqnA1Amp}\end{equation}
	
	and the solution for $A_1$ and $A_2$ yields
	
	\begin{equation}
		\begin{aligned}
			A_1=a_1e^{\alpha x}+b_1e^{-\alpha x}\\
			A_2=-\mathrm{j}\sqrt{\frac{m_1k_1\rho_1}{m_2k_2\rho_2}}(a_1e^{\alpha x}-b_1e^{-\alpha x})
		\end{aligned}
		\label{eq:TwoAmplitudesAmp}\end{equation}
	
	where $\alpha=\frac{\sqrt{m_1m_2k_1k_2}}{4}$. Applying the same boundary conditions as equation (\ref{eq:TwoBCsv}), we get the complete solution of equation(\ref{eq:WaveEqnRho}) as
	
	\begin{equation}
		\begin{aligned}
			v_1=a\frac{e^{\alpha x}+e^{-\alpha x}}{2}e^{\mathrm{j}(\omega_1t-k_1x+\theta)}\\
			v_2=\sqrt{\frac{m_1k_1\rho_1}{m_2k_2\rho_2}}a\frac{e^{\alpha x}-e^{-\alpha x}}{2}e^{-\mathrm{j}(\omega_2t-k_2x-\theta+\frac{\pi}{2})}
		\end{aligned}
		\label{eq:SolutionvAmp}\end{equation}
	
	Now, instead of periodically varying, the amplitudes of both waves are growing exponentially. This way we can get a piece of "gain" material for both frequencies. A special case is that when $\omega_1=\omega_2$, both waves are of the same frequency. In this case, there will be two waves at the output end: one is the amplified original wave, and the other one is the generated wave with the same frequency but with conjugated phase.
	
	Here we showed two possibilities enabled by space-time modulation as described in Eqns.~(\ref{eq:AssumeConversion}) and ~(\ref{eq:AssumeAmp}), namely unidirectional frequency conversion and parametric amplification, which cannot be easily realized in reciprocal systems. However, we would like to emphasize that these are just two classes of solutions of a space-time modulated system, and there are much more possibilities depending on the modulation strategy.
	
	\subsection{Sound propagation in media with space-time modulated bulk modulus}
	
	Similar to the case with modulated density, a waveguide with space-time modulated bulk modulus can also produce non-reciprocal one-way wave behaviors such as frequency conversion, amplification and phase conjugation. We will show that the solutions will have similar structure but different scaling constants.. The wave equation in such a medium is written as
	
	\begin{equation}
		\begin{aligned}
			-\frac{\partial p}{\partial x}=\rho\frac{\partial v}{\partial t}\\
			-\frac{\partial (B p)}{\partial t}=\frac{\partial v}{\partial x}
		\end{aligned}
		\label{eq:NewtonHooke2}\end{equation}
	
	where $B=\frac{1}{\kappa}$ is the compressibility (or effective compressibility) of the medium under study. With these two equations, the wave equation can be obtained in the form of
	
	\begin{equation}
		\frac{\partial^2p}{\partial x^2}=\rho\frac{\partial^2B}{\partial t^2}p+2\rho\frac{\partial B}{\partial t}\frac{\partial p}{\partial t}+\rho B\frac{\partial^2p}{\partial t^2}
		\label{eq:WaveEqnMod}
	\end{equation}
	
	Suppose the compressibility of the medium in the waveguide is varying with the form
	
	\begin{equation}
		B=B_1[1+m_1\cos(\Omega t-\beta x)]
		\label{eq:BChange1}\end{equation}
	for the wave with frequency $\omega_1$, and 
	\begin{equation}
		B=B_2[1+m_2\cos(\Omega t-\beta x)]
		\label{eq:BChange2}\end{equation}
	for the wave with frequency $\omega_2$, and the solution has the form of $p=p_1+p_2$, where
	
	\begin{equation}
		\begin{aligned}
			p_1=A_1(x)e^{\mathrm{j}(\omega_1t-k_1x)}\\
			p_2=A_2(x)e^{\mathrm{j}(\omega_2t-k_2x)}
		\end{aligned}
		\label{eq:TwoWavesp}\end{equation}.
	
	For frequency conversion we can apply the modulation in the form  the same as equation (10-11). Putting it in equation (\ref{eq:WaveEqnMod}) and equate the terms with the same frequency and wave number, we can get the same differential equation for $A_1$ as equation (\ref{eq:EqnA1}). Solving the $A_1$ and $A_2$ we get
	
	\begin{equation}
		\begin{aligned}
			A_1=a_1e^{\mathrm{j}\alpha x}+b_1e^{-\mathrm{j}\alpha x}\\
			A_2=-\frac{B_1}{B_2}\sqrt{\frac{m_1k_2}{m_2k_1}}(a_1e^{\mathrm{j}\alpha x}-b_1e^{-\mathrm{j}\alpha x})
		\end{aligned}
		\label{eq:FormA12}\end{equation}
	
	Applying the boundary condition at $x=0$: 
	
	\begin{equation}
		\begin{aligned}
			p_1(0,t)=a e^{\mathrm{j}(\omega_1t+\theta)}\\
			p_2(0,t)=0
		\end{aligned}
		\label{eq:TwoBCsp}\end{equation}
	
	the following complete solution can be found:
	
	\begin{equation}
		\begin{aligned}
			p_1=a\cos(\alpha x)e^{\mathrm{j}(\omega_1t-k_1x+\theta)}\\
			p_2=-\frac{B_1}{B_2}\sqrt{\frac{m_1k_2}{m_2k_1}}a\sin(\alpha x)e^{\mathrm{j}(\omega_2t-k_2x+\theta+\frac{\pi}{2})}
		\end{aligned}
		\label{eq:TwoAmplitudes}\end{equation}
	
	Similarly, if the applied modulation is the same as equation (\ref{eq:AssumeAmp}), parametric amplification and phase conjugation can be achieved.

	\section{Metamaterial realization of space-time modulated acoustic media}
	
	In some cases, for elastic waves in solids, material properties can be tuned by an external field. For example, one can change the elastic modulus by applying a voltage to a piece of piezoelectric material. However, directly changing the properties of a fluid is challenging since modulating density or bulk modulus usually means modulating the temperature in a fast and controlled manor. In recent years, the development of the concept of metamaterials has enabled theoretically arbitrary values of effective density or modulus by carefully designing the subwavelength structures of the material \cite{cummer2016controlling,ma2016acoustic}. 
	
	By introducing active elements into the metamaterials, the achieved values can further be controlled. It opens up the possibility to manipulate the effective parameters spatiotemporally without too much cost to achieve enough modulation depth at the desired rate. Piezoelectric membranes, for example, have been proposed for tuning the effective density to change the working frequency \cite{ma2014acoustic,xiao2015active}. However, such membranes can only achieve limited tunability at the cost of extremely high voltages, and it is hard to modulate dynamically. Here we propose the dynamic change of a system with an array of side-loaded Helmholtz resonators, in which the effective bulk modulus can be modulated through small motion of the back walls, as shown in Fig. 2.
	
	\begin{figure}
		\includegraphics[width=0.9\linewidth]{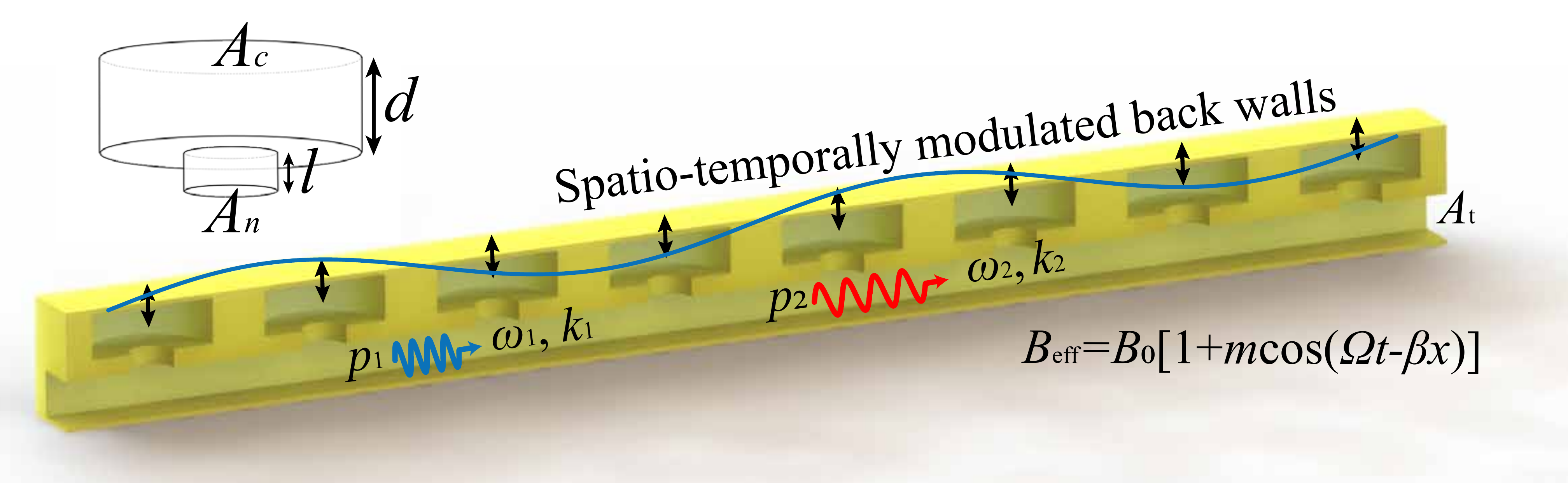}
		\caption{One proposed realization of a space-time modulated acoustic wave system. The moving back wall allows the cavity volume to be modulated. The inset shows the dimensions of a single Helmholtz resonator.} 
		\label{fig:Fig2}\end{figure}
	
	The effective compressibility for a metamaterial system composed of static resonators can be written as \cite{fang2006ultrasonic,lee2009acoustic}:
	\begin{equation}
		B_{eff}(\omega)=B_0[1+\frac{F\omega_0^2}{\omega_0^2-\omega^2+\rm{j}\omega\gamma}]
		\label{eq:OldEffCompressStatic}\end{equation}
	where $F=n^2A_n^2/A_t$ is the geometrical factor, $n$ is the number of cells per meter, $A_n$ is the cross sectional area of the neck, and $A_t$ is the cross sectional area of the waveguide.  The factor $\gamma$ is associated with the loss in the system. The parameter $\omega_0=\frac{c_0^2A_n}{A_cdl}$ is the resonant angular frequency, where $A_c$ is the area of the cavity back wall, $d$ is the height of the cavity, and $l$ is the corrected length of the neck. 
	
	We now consider a metamaterial in which the back wall of the cavity is moving sinusoidally so that $d$ is changing with time as $d=d+\delta d\rm{cos}(\Omega t-\beta x)$.  We can rewrite the effective compressibility as 
	\begin{equation}
		B_{eff}(\omega,d)=B_0+\frac{\frac{nA_n}{A_tl\rho_0}}{\frac{c_0^2A_n}{A_tld}-\omega^2+\rm{j}\omega\gamma}
		\label{eq:EffCompress}\end{equation}
	Therefore, the modulation depth $m$ at a given frequency can be estimated with
	\begin{equation}
		m(\omega)=\frac{|\rm{Re}[B_{eff}(\omega,d+\delta d)]-\rm{Re}[B_{eff}(\omega,d-\delta d)]|}{2\rm{Re}[B_{eff}(\omega,d)]}.
		\label{eq:ModulationDepth}\end{equation}
	The wavenumber for the system can be calculated with $k(\omega)=\omega \rm{Re}[\sqrt{B_{eff}(\omega)\rho_0}]$. Hence, for a given system and the targeted frequencies, all the modulation parameters can be theoretically calculated to achieve different functionalities such as frequency conversion or parametric amplification. Also, the rate of conversion and amplification can be estimated by calculating $\alpha$ in Eqns.(\ref{eq:FormA1},\ref{eq:FormA2}) and Eqn. (\ref{eq:TwoAmplitudesAmp}).
	
	\section{Numerical Simulations}
	
	The analytical model is verified here with 1D FDTD effective medium simulation. The background media is lossless air with density $\rho_0=1.2\rm{kg/m^3}$ and speed of sound $c_0=343\rm{m/s}$. The time step is $2\times10^{-6}$s and the grid is $1\times10^{-3}$m. In the simulation, we study the wave propagation in an effective medium approximation of the metamaterial structure with modulated cavity height, as described in Section III. The number of resonators is $n=25 m^{-1}$. The round necks have a radius of 5 mm and corrected length $l=6\rm{mm}$, and the cavity is a cylinder with radius of radius of 14 mm and height $d=4\rm{mm}$ so that the resonant frequency is 3980 Hz. The modulation depth is $\delta d=0.4$ mm. The wave propagates in a non-modulated medium before (upstream) and after (downstream) the modulated section. A sinusoidal wave is incident from the upstream direction.  To show that the wave interacts differently with the space-time modulation in different directions, we simulate the wave incident from both directions. The backward incident wave was simulated by switching the sign of modulation wave number to $-\beta$.
	\begin{figure}
		\includegraphics[width=0.8\linewidth]{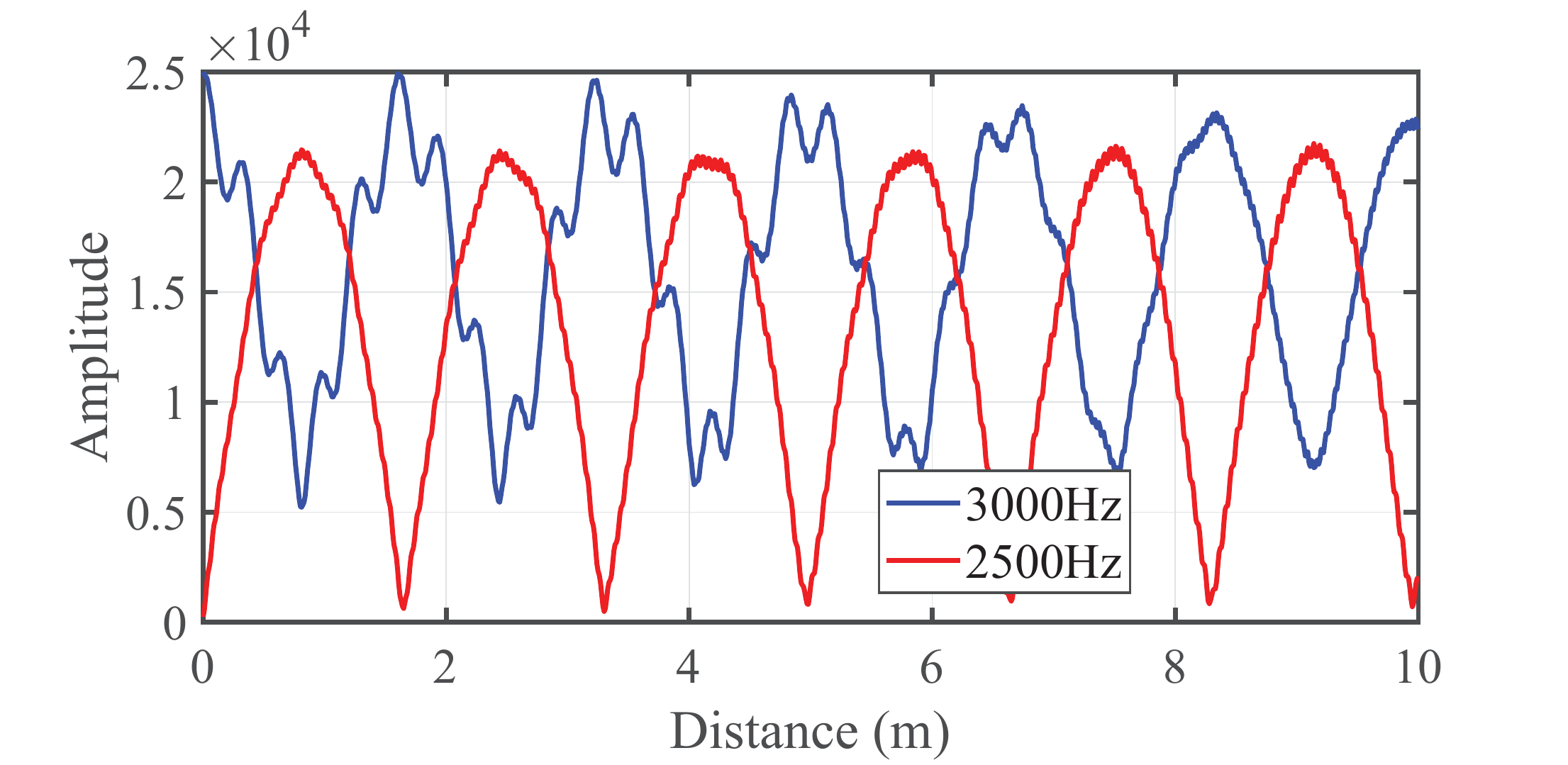}
		\caption{Two targeted frequency components along the space-time modulated media in the FDTD simulation. The wave energy transfers back and forth between the two components.} 
		\label{fig:Fig3}\end{figure}
	
	We first simulate the wave propagation in a modulated medium where the modulation satisfies Eqn. (\ref{eq:AssumeConversion}), so that the device acts as a frequency converter. The target input and output frequencies are $f_1=3000\rm{Hz}$ and $f_2=2500\rm{Hz}$, respectively. The modulated length is 10 m, with upstream length 0.3 m and downstream length 30 m so that the reflection from the boundary can be easily time-gated. To study the steady state wave behavior, the data of first 80 ms is excluded from processing. The simulated pressure field in the modulated region is recorded for 100 ms after the signals become stable. Fig. 3 shows the amplitude of the two frequency components along the modulated medium. From the figure it is clearly seen that the wave energy gradually shifts from $f_1$ to $f_2$ and then transfer back to $f_1$, as predicted in the theory. The distance for $f_2$ to get to its first peak is 0.81 m in the simulation, while the corresponding theoretical calculation is 0.78 m, in good agreement. The small discrepancy originates from the finite resolution in the simulation.
	
	An advantage of the proposed metamaterial implementation of an acoustic space-time modulated medium lies in that all the different functionalities can be achieved without changing the structure of the system, and the behavior depends only on how the resonators are modulated. For the functionalities described in Section II, the corresponding simulation results are summarized in Fig. 4. For frequency converter, the target input and output frequencies are $f_1=3000\rm{Hz}$ and $f_2=2500\rm{Hz}$, respectively. The signal on the incident side and transmission side of the modulated medium was recorded for analysis. Fig. 4(a) and Fig. 4(b) show the spectrum of the incident wave and transmitted wave when the wave incidents from positive direction and negative direction, respectively. From the figures we can see that for the wave incident from the positive direction, the main frequency component is effectively converted from 3000 Hz to 2500 Hz. The wave incident from the negative direction is not significantly affected by the modulation.
	
	\begin{figure}
		\includegraphics[width=0.9\linewidth]{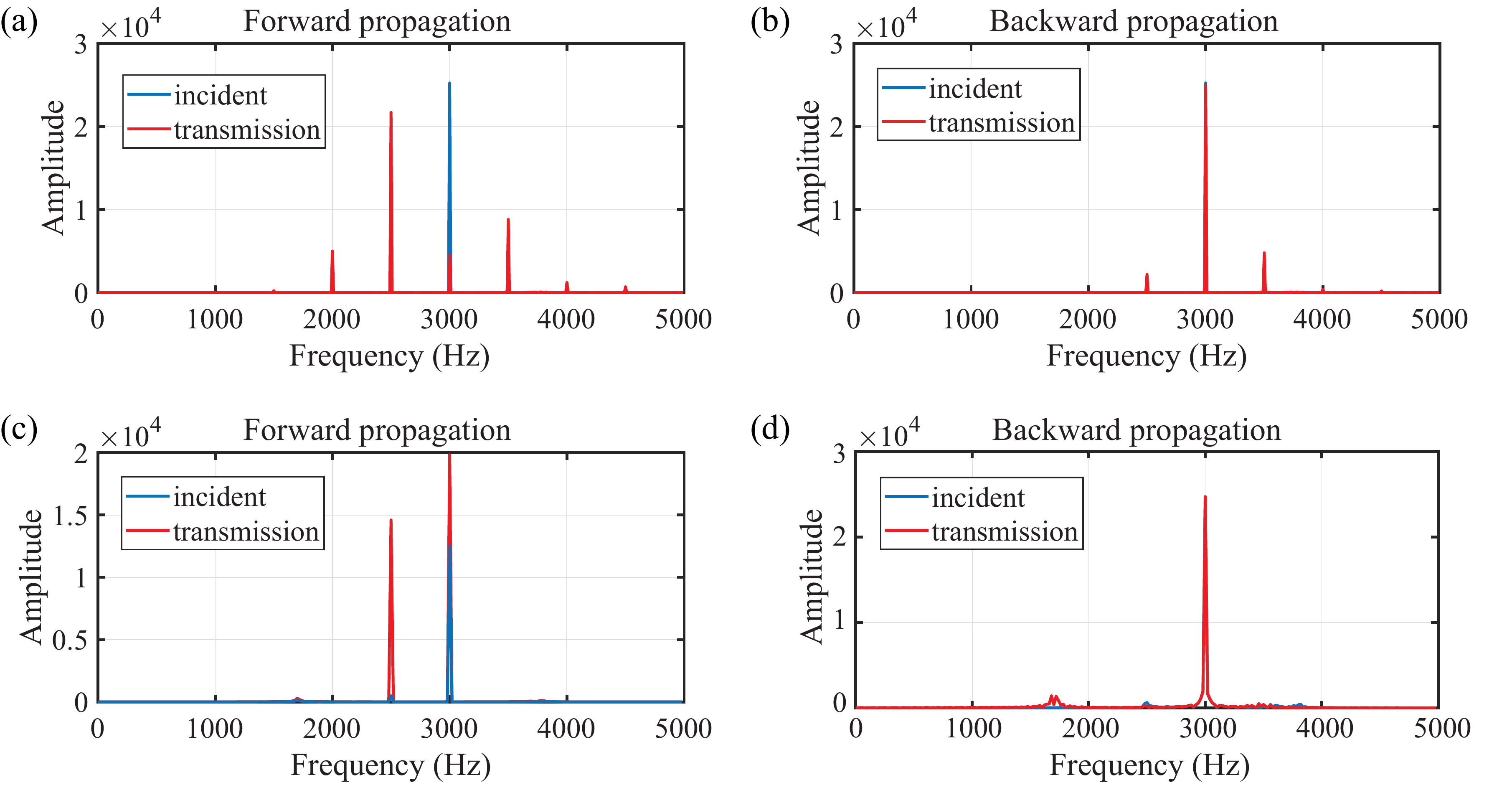}
		\caption{Simulated spectrum of the incident and transmitted waves when the device is modulated as a frequency converter or parametric amplifier for both directions. (a) Frequency converter, the wave is converted from $f_1$ to $f_2$ while propagating in the positive direction. (b) when it's propagating in the negative direction, it doesn't interact with the modulation so that the transmitted is almost the same as incident wave. (c) Parametric amplifier, $f_1$ got amplified in the positive direction, generating $f_2$ in the mean time. (d) The wave is not affected while propagating in the negative direction.} 
		\label{fig:Fig4}\end{figure}
	
	For non-reciprocal one-way wave transmission at 3000 Hz, the isolation level measures the efficiency of one-way isolation of the acoustic waves, defined as the contrast between the transmitted waves when the wave incidents from opposite directions. Although the isolation level can theoretically reach infinity, in the simulation, it reaches 15 dB. This is due to the incomplete conversion between two waves, and conversion to other frequency components. If we look at 2500 Hz, then the isolation level reaches 20 dB in simulation. Counterintuitively, the isolation level can be further increased using smaller modulation depth which better approximates the weak modulation assumption in our derivation. For example, when modulation depth $\delta d=0.04$ mm is applied, the isolation level for 3000 Hz and 2500 Hz reaches 22 dB and 46 dB, respectively. Other possible ways for improvement are further discussed in Section VI. 
	
	If the device is modulated according to the space-time profile of Eqn.~(\ref{eq:AssumeAmp}), it acts as a parametric amplifier. Both $f_1$ (3000 Hz) and $f_2$ (2500Hz) are growing exponentially. The simulated spectrum on the incident side and transmission side is shown in Fig. 4(c). In the simulation, the modulated length is 0.5 m to prevent the signals from growing too large. Upstream length is 5 m while downstream length is 20 m to prevent reflection. The modulation depth remains unchanged. From Fig. 4(c) it is seen that as the incident wave propagates in the space-time modulated media, the waves get amplified while generating the other frequency component. The growth rate $\alpha$ can be calculated by taking the amplitude ratio between the incident $f_1$ and the generated $f_2$, and then solve Eqn. (\ref{eq:SolutionvAmp}). The calculated growth rate from simulated result and theoretical calculation are $\alpha=1.9917$ and $\alpha=2.0082$, respectively, which again shows excellent agreement between the theory and simulation. Fig. (d) shows the wave propagating in the negative direction. We can see that the wave is not affected by the modulation. Therefore, the device facilitates one-way parametric amplification.

	\section{Summary and Conclusions}
	Here we have developed a theory to characterize the waves propagating in an acoustic space-time modulated medium. Specifically, when the medium is modulated such that different modes can be coupled through that modulation, non-reciprocal functionalities such as one-way frequency conversion and parametric amplification can be achieved, which is beyond the reach of linear time-invariant systems. We show how such a medium could be implemented using small structural modulation of Helmholtz resonators, and a numerical FDTD simulation based on such a design is developed to show that the theory is valid and that the predicted behavior can be delivered in practice. The simulation results showed excellent agreement with theoretical calculations.
	
	\begin{figure}
		\includegraphics[width=0.95\linewidth]{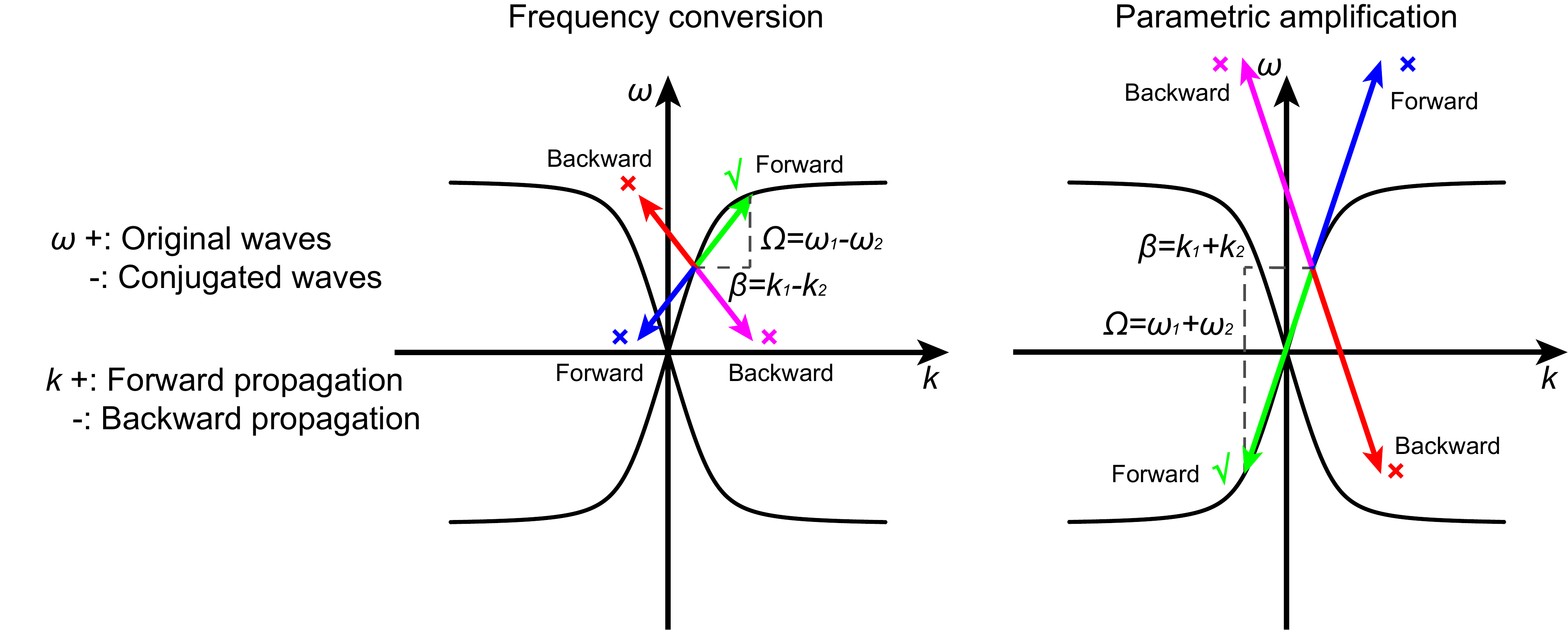}
		\caption{Modulation facilitates mode hopping on the $\omega-k$ diagram. Only supported modes of the system can be coupled through modulation. This feature allows one-way manipulation of waves.} 
		\label{fig:Fig5}\end{figure}
	
	To understand the physical mechanism of the proposed system, the corresponding one-way phenomena can be viewed as mode conversion on the $\omega-k$ diagram, which corresponds to the inter-band and intra-band photon transition in optical systems, as shown in Fig. 5. It is shown that the two otherwise orthogonal modes are coupled through the modulation. However, the wave will not interact with the modulation if the modulation does not lead to any allowed mode in the system. Therefore, the prescribed modulation only works for forward propagating wave, as is shown in Fig. 5. For each type of modulation, the green and blue arrows denote the modulation when the wave travels in the forward direction, while the red and magenta arrows denote the modulation when the wave travels in the backward direction. In the diagram it is seen that only the designed mode is coupled through modulation. This is analogous to the mechanism of modulated phononic crystal in the mass-spring systems. 
	
	Since the modulation relies on shifting on the $\omega-k$ diagram, it is important to avoid coupling to unwanted modes for efficient spectral manipulation. For example, if the material has a purely linear $\omega-k$ diagram, the modes with $\omega=\omega_1\pm n\Omega$ will all be coupled, and all of these couplings needs to be taken into consideration in the theory, which is beyond the scope of this paper. 
	
	Our work outlines a robust and efficient means of versatile manipulation of an incident wave compared with non-linear devices which can only generate harmonics of the fundamental mode and there is little control over the ratio between fundamental mode and its harmonics. Also, since the device does not require operation near resonant frequencies, it is less sensitive to losses and fabrication errors, which provides a significant advantage in realization compared with non-reciprocal devices based on coupled resonances. We have also simulated the system with embedded loss by assigning $\gamma=0.01$ in Eqn.~(\ref{eq:OldEffCompressStatic}) and the results remain essentially unchanged. This indicates that internally dissipated energy is compensated for by the modulation. This feature makes the whole system more robust against losses in real implementations.
	
	The mode conversion process mediated by the frequency and momentum of the modulation, is an 'indirect phononic transition', in analogy with indirect electronic transitions in semiconductors, where interaction with optical signals and phonons changes the energy and momentum of electrons. Here we offered a new perspective on solving space-time equations directly instead of analyzing band structures with Floquet theory. It provides more detailed information about how do waves change gradually in such systems.  These non-reciprocal phenomena open many possibilities for unprecedented wave control capability. For example, mode conversion enables advanced spectrum control for one-way and encoded communications, directional energy transmission control, and directional band gap for acoustic rectifiers and topological insulators; parametric amplification provides a possible way for designing the gain media in parity-time (P-T) symmetric systems and powerful acoustic radiator designs; phase conjugation enables all-angle retroreflectors, acoustic lasing, mathematical operation for data processing, and may find applications in acoustic communication systems.

	\section{\label{sec:citeref}References}
	\bibliographystyle{apsrev4-1}
	\bibliography{references}

\end{document}